\documentclass[11pt,a4paper,english,twoside]{article}

\usepackage{a4wide}
\usepackage{amssymb, amsmath}
\usepackage{graphicx}
\usepackage[all]{xy}
\usepackage{hyperref}
\hypersetup{linktocpage}
\usepackage{enumerate}
\usepackage{xcolor}
\usepackage{dsfont}
\usepackage{empheq}
\usepackage{cite}
\usepackage{float}

\newcommand{\beq}{\begin{equation}}
\newcommand{\eeq}{\end{equation}}
\def\bea#1\eea{\begin{align}#1\end{align}}
\def\beal#1\eeal{\begin{subequations}\begin{align}#1\end{align}\end{subequations}}
\newcommand{\nn}{\nonumber}
\newcommand{\w}{\wedge}
\renewcommand{\i}{\ensuremath{\textnormal{i}}}

\newcommand{\eq}[1]{\begin{equation}#1\end{equation}}
\newcommand{\spl}[1]{\begin{split}#1\end{split}}
\newcommand{\al}[1]{\begin{align}#1\end{align}}

\def\del {\partial}
\def\d {{\rm d}}
\def\mmm {\mathcal{M}}
\def\f {{\rm \texttt{f}}}

\def\ep {\epsilon}

\begin{document}
\numberwithin{equation}{section}

\begin{titlepage}

\begin{flushright}
CERN-TH-2018-137
\end{flushright}

\begin{center}

\phantom{DRAFT}

\vspace{2.8cm}

{\LARGE \bf{Laplacian spectrum on a nilmanifold,\vspace{0.3cm}\\truncations and effective theories}}\\

\vspace{2 cm} {\Large David Andriot$^{1}$ and Dimitrios Tsimpis$^{2}$}\\
 \vspace{0.9 cm} {\small\slshape $^1$ Theoretical Physics Department, CERN\\
1211 Geneva 23, Switzerland}\\
 \vspace{0.2 cm} {\small\slshape $^2$ Universit\'e Lyon 1, CNRS/IN2P3, IPNL\\
69622 Villeurbanne, France}\\
\vspace{0.5cm} {\upshape\ttfamily david.andriot@cern.ch; tsimpis@ipnl.in2p3.fr}\\

\vspace{3cm}

{\bf Abstract}
\vspace{0.1cm}
\end{center}

\begin{quotation}
\noindent
Motivated by low energy effective theories arising from compactification on curved manifolds, we determine the complete spectrum of the Laplacian operator on the three-dimensional Heisenberg nilmanifold. We first use the result to construct a finite set of forms leading to an ${\cal N}=2$ gauged supergravity, upon reduction on manifolds with SU(3) structure. Secondly, we show that in a certain geometrical limit the spectrum is truncated to the light modes, which turn out to be left-invariant forms of the nilmanifold. We also study the behavior of the towers of modes at different points in field space, in connection with the refined swampland distance conjecture.

\end{quotation}

\end{titlepage}

\newpage

\noindent\rule[1ex]{\textwidth}{1pt}
\vspace{-0.8cm}

\tableofcontents

\vspace{0.5cm}
\noindent\rule[1ex]{\textwidth}{1pt}
\vspace{0.3cm}

\section{Introduction}

String compactifications is the main framework for string phenomenology, whereby one considers a four-dimensional maximally-symmetric space-time together with six extra space dimensions forming a compact manifold $\mmm$. The four-dimensional theory obtained after dimensional reduction is crucially dependent on the geometry of $\mmm$ and the background fields living on it. While the main focus of string phenomenology involves Calabi-Yau manifolds, i.e.~the case where $\mmm$ admits a Ricci-flat metric, curved manifolds (which we take to mean manifolds which are not Ricci-flat, e.g.~most group manifolds or cosets) present a number of phenomenologically appealing characteristics. String backgrounds with curved $\mmm$ may widely populate the string landscape away from the lamppost, providing new phenomenological effects worth being understood. For instance, the curvature of $\mmm$ is known to generate a potential for specific four-dimensional scalar fields, thus providing them with a mass, otherwise difficult to generate classically. Another example comes from the fact that classical de Sitter solutions (at least with parallel orientifold planes) require a negative Ricci scalar for $\mmm$ (see e.g.~\cite{Haque:2008jz, Andriot:2016xvq}). Negatively-curved manifolds are in a certain mathematical sense much more numerous than positively-curved ones \cite{berger}. They are however less well-studied in the context of compactification than their positively-curved counterparts. Here we will focus on nilmanifolds, group manifolds based on nilpotent Lie algebras, which are a special case of negatively-curved manifolds. They offer a non-trivial yet tractable playground on which exact calculations can be performed.

Obtaining a four-dimensional low energy effective theory from a ten-dimensional one requires knowledge of the spectrum and eigenmodes of the Laplacian operator on $\mmm$. Indeed, the eigenvalues correspond to masses of four-dimensional states for a free theory, giving access to energy hierarchies, while the eigenmodes provide a basis to expand the ten-dimensional field, prior to the dimensional reduction; this procedure is discussed in more detail in Section \ref{sec:effth}. Harmonic analysis on nilmanifolds, in particular on the three-dimensional Heisenberg manifold $M$,\footnote{We use $\mmm$ for the six-dimensional internal manifold and $M$ for the three-dimensional Heisenberg manifold.} has been considered before in the mathematical literature \cite{thang,Gordon, Gornet1, Gornet2, Muller}, however the results are usually not presented in a way familiar to most physicists. Moreover the analyses available typically do not consider the dependence of the spectrum on the metric moduli -- a rather useful piece of information from the physics point of view as it directly affects the masses of the physical fields.

In \cite{Andriot:2016rdd} we determined the scalar spectrum of the Laplacian on $M$, and its dependence on the metric moduli. The study revealed some potentially promising phenomenological applications and allowed us to test the accuracy of available codes for the numerical determination of the spectrum. Ultimately, a complete analysis of the mass spectrum and its implications for four-dimensional physics requires the knowledge of the
Laplacian spectrum for all differential forms on the manifold.\footnote{The spectrum of the Laplacian on a different nilmanifold has been worked-out in \cite{Camara:2009xy}. Some similarities can be found in the eigenvalues of the spectrum, although they do not match the ones found here.} This is the main subject of the present paper.

Phenomenological implications of our results are beyond the scope of this work, and will be examined elsewhere \cite{wip}, nevertheless the knowledge of the Laplacian spectrum already allows us to make interesting observations concerning effective actions and reductions on manifolds with SU(3) structure, as considered in the literature \cite{Gurrieri:2002wz, DAuria:2004kwe, House:2005yc, Grana:2005ny, Louis:2006kb, KashaniPoor:2006si, Cassani:2009ck}.\footnote{Here by the term ``effective action''  or ``reduction'' we understand neither a low-energy effective action (LEEA) nor a consistent truncation (CT). By a LEEA we mean a truncation of the ten-dimensional theory to a finite subset of four-dimensional fields that correctly describes the four-dimensional physics below a certain energy scale. A CT is a four-dimensional theory such that all its solutions lift to solutions of the ten-dimensional theory. In general the notions of CT and LEEA are completely distinct in the sense that a CT is not a LEEA and vice versa.} The strategy adopted in those references is to postulate the existence of a finite set of forms on $\mmm$ satisfying a list of constraints. A truncation ansatz is then defined whereby the ten-dimensional fields are expanded on this set of forms. Plugging the truncation ansatz in the ten-dimensional action can then be seen to give rise to a four-dimensional theory, more precisely an ${\cal N}=2$ gauged supergravity. This non-trivial result suggests that these effective actions may be CT of ten-dimensional supergravity, although this has not be proven to date. It is also unclear whether these effective actions can be thought of as LEEA.

Thanks to our basis of Laplacian eigenforms, we will be able to construct an explicit example of the finite set of forms considered in these reductions on manifolds with SU(3) structure. We will do so for two cases: $\mmm=M \times M$, or $\mmm=M \times T^3$. The finite set of forms will satisfy all required constraints, giving rise to an ${\cal N}=2$ gauged supergravity. We also discuss whether the reduction on our set of forms may be thought of as a LEEA, although a definitive answer must in any case await a complete Kaluza--Klein analysis of the mass spectrum around an appropriate ten-dimensional (flux) solution.

Finally we show that, in a certain geometrical limit, the Laplacian spectrum admits an interesting low energy truncation to massless and massive light modes, which turn out to be  Maurer--Cartan (left-invariant) forms. We also analyse the behavior of the towers of modes in different regimes, to make contact with the recent discussion on the so-called ``refined swampland distance conjecture'' \cite{Ooguri:2006in, Klaewer:2016kiy}. This conjecture states that for effective theories of quantum gravity, a tower of states becomes light when moving sufficiently far in field space, thus spoiling an initial effective description; a more detailed presentation is given in Section \ref{sec:swamp}. Interestingly,  this behavior is {\it not} necessarily what we find for the towers of the eigenmodes of the Laplacian, although we are missing a piece of information that would allow us to conclusively decide whether or not there is a tension with the conjecture.

The outline of the paper is as follows. In Section \ref{sec:lapl} we determine the spectrum and eigenforms of the Laplacian operator on the three-dimensional Heisenberg nilmanifold. In light of these results we examine in Section \ref{sec:truncation} the truncation to the constrained finite set of forms considered in reductions on manifolds with SU(3) structure. In Section \ref{sec:lowenergy} we define a certain geometrical limit in which the truncation to the light (not necessarily massless) eigenmodes of the Laplacian coincides with Maurer--Cartan forms. Further limits are studied in connection with  the swampland discussion. We conclude with a discussion of our results in Section \ref{sec:effth}.

\section{Laplacian spectrum}\label{sec:lapl}

\subsection{Nilmanifold geometry and scalar spectrum}\label{sec:geoscalar}

We consider the three-dimensional nilmanifold $M$ built from the nilpotent Heisenberg algebra
\eq{\label{heis}[V_1,V_2]=- \f\, V_3~,~~~[V_1,V_3]=[V_2,V_3]=0~,}
with structure constant $\f=-f^3{}_{12}$. The Maurer--Cartan one-forms $e^{a=1,2,3}$, dual to the vectors $V_a$, satisfy
\eq{\label{mcheis}
\d e^3= \f\, e^1\wedge e^2~;~~~\d e^1=0~;~~~\d e^2=0
~.}
These vectors and one-forms provide a basis of the (co)-tangent space of the group manifold $M$. We choose angular coordinates $x^{m=1,2,3} \in [0,1]$, constant (positive) radii $r^{m=1,2,3}$, and the following parametrisation
\eq{\label{a}e^1=r^1\d x^1~;~~~e^2=r^2\d x^2~;~~~e^3=r^3\left(\d x^3+N x^1\d x^2\right)~;~~~N=\frac{r^1r^2}{r^3}\f \in\mathbb{Z}^*~.}
One defines equivalently the vielbein $e^a{}_m$ and its inverse as $e^a = e^a{}_m \d x^m, \ V_a = e^m{}_a \del_m$. In terms of physical dimension, the coordinates $x^m$ are dimensionless while $r^m$ and $e^a$ have the dimension of a length, and $\f$ that of the inverse of a length. The manifold $M$ is compact thanks to the following discrete identifications
\eq{\label{heisids}
x^1\sim x^1+n^1~;~~~x^2\sim x^2+n^2~;~~~x^3\sim x^3+n^3- n^1N x^2~;~~~n^1,n^2,n^3\in\{ 0,1\}~.
}
They correspond to the lattice action, making $M$ the quotient of a nilpotent group by a discrete subgroup, i.e.~a nilmanifold. The one-forms $e^a$ are invariant under \eqref{heisids}, thus globally defined. Geometrically, the discrete identifications \eqref{heisids} indicate that $M$ is a twisted $S^1$ fibration over $T^2$, i.e.~a twisted torus, with fiber coordinate $x^3$ and a base parameterised by $x^1, x^2$. As we showed in \cite{Andriot:2016rdd} the most general metric on this manifold is parameterised as
\eq{\label{expl}\d s^2=\big(e^1+a e^3\big)^2+\big(e^2+b e^3\big)^2+\big(e^3\big)^2~,~~~a,b\in\mathbb{R}~,}
where the parameter $c$ of \cite{Andriot:2016rdd} can be set to $1$ without loss of generality. One deduces $\sqrt{g}= r^1 r^2 r^3$, and the volume is given by
\eq{\label{vol} V=\int\d^3 x\sqrt{g}= r^1 r^2 r^3~.}
In this paper, we restrict ourselves to the case where $a=b=0$ and use the metric $\d s^2= \delta_{ab} e^a e^b$.

We found in \cite{Andriot:2016rdd} the eigenmodes and eigenvalues of the Laplacian operator $\Delta$ acting on a scalar field $\varphi$
\eq{\label{Lap} \Delta \varphi = \nabla^2\varphi=\frac{1}{\sqrt{g}}
\partial_m\left(\sqrt{g}g^{mn}\partial_n\varphi\right)
~.}
In the case of the nilmanifold $M$, one also has $\Delta \varphi = \delta^{ab} V_a V_b\, \varphi$ \cite{Andriot:2016rdd}. The spectrum was obtained for the metric \eqref{expl}; including parameters $a,b$ complicates the expressions and does not add physical states nor leads to major qualitative changes, so we restrict here to $a=b=0$. Two sets of orthonormal eigenfunctions were found: $v_{p,q}$, independent of $x^3$, and $u_{k,l,n}$, dependent on $x^3$, verifying
\beq
\big(\Delta+\mu^2_{p,q}\big) v_{p,q}=0 \ ,\quad \big(\Delta+M^2_{k,l,n}\big) u_{k,l,n}=0 \ . \label{Lapeq}
\eeq
The spectrum is given by
\eq{\label{masses}\spl{
\mu_{p,q}^2 &= p^2 \, \left(\frac{2\pi}{r^1}\right)^2 + q^2\, \left(\frac{2\pi}{r^2}\right)^2 \ ,\\
M^2_{k,l,n} &= k^2 \, \left(\frac{2\pi }{r^3}\right)^2 +(2n+1) |k|\, \frac{2\pi |\f|}{r^3 } \ ,
}}
while the orthonormal modes are
\eq{\label{uv}\spl{
v_{p,q}(x^1,x^2)&= \frac{1}{\sqrt{V}}\, e^{2\pi \i p x^1 } e^{2\pi \i q x^2 } \ ,\\
u_{k,l,n}(x^1,x^2,x^3)&= \sqrt{\frac{r^2}{|N|V}}\frac{1}{\sqrt{2^n n! \sqrt{\pi}}}\,  e^{2\pi \i k (x^3+N\, x^1x^2)} e^{2\pi \i l x^1}\sum_{m\in\mathbb{Z}} e^{2\pi \i k m x^1} \Phi^{\lambda}_n(w_m) \ ,
}}
with $\lambda= k \frac{2\pi \f}{r^3}$ and $w_m=r^2 \left( x^2 + \frac{m}{N} + \frac{l}{kN} \right) $. The integers have the following ranges
\beq
p,q\in\mathbb{Z} \quad {\rm and}\quad k\in\mathbb{Z}^*~,~n\in\mathbb{N}~,~l=0,\dots, |k|-1\ .
\eeq
The function $\Phi^{\lambda}_n$ is defined for $\lambda\in\mathbb{R}^*$ in terms of the normalised Hermite functions
\eq{\label{md}
\Phi^{\lambda}_n(z)=|\lambda|^{\frac14}\, \Phi_n(|\lambda|^{\frac12}z)\ , \quad \Phi_n(z)=e^{-\frac12 z^2}H_n(z)
\ ,\quad n\in\mathbb{N}~,}
where $H_n$ are the Hermite polynomials defined as $H_n(y)=(-1)^n e^{y^2} \del^n_y e^{-y^2}$.

\subsection{One-form spectrum}

\subsubsection{Hodge decomposition and exact one-forms}

We are interested in the one-forms $B$ on $M$ satisfying the eigenvalue equation
\eq{
\Delta B = \Upsilon B \ ,\label{equation}
}
for a constant $\Upsilon$. The Laplacian operator is given by $\Delta B=(*\, \d*\d + \d*\d\, *)B$ where $\d$ is the exterior derivative on $M$ with the Hodge star $*$. The latter is defined on a $p$-form in $D$ dimensions as
\eq{\label{Hodgestar}\spl{
& *(\d x^{m_1} \w \dots \w \d x^{m_p}) = \frac{\sqrt{|g|}}{(D-p)!} \epsilon^{m_1 \dots m_p}{}_{n_{p+1} \dots n_D}\ \d x^{n_{p+1}} \w \dots \w \d x^{n_D} \\
\Leftrightarrow \ & *(e^{a_1} \w \dots \w e^{a_p}) = \frac{1}{(D-p)!} \epsilon^{a_1 \dots a_p}{}_{a_{p+1} \dots a_D}\ e^{a_{p+1}} \w \dots \w e^{a_D}  \ ,
}}
with $\epsilon_{1 \dots d} = 1$ and curved or flat indices raised by $g_{mn}$ or $\delta_{ab}$ respectively. We also recall for a $p$-form $A_p$ that $*^2 A_p = s (-1)^{p(D-p)} A_p = s (-1)^{p(D+1)} A_p$ where $s$ is the signature of the $D$-dimensional space, i.e. $*^2=1$ for our $M$.

The Hodge decomposition of $B$ gives
\eq{\label{hodge}B=\d\varphi+\d^{\dagger}b_2+h~,}
where $\varphi$ is a globally-defined scalar, $b_2$ a globally-defined two-form and $h$ a globally-defined harmonic one-form on the nilmanifold. The three terms on the right-hand side above are orthogonal to each other with respect to the canonical pairing of one-forms on $M$. Let us examine the first one and look for exact one-forms solving the eigenvalue equation. The function $\varphi$ can be expanded on the basis of eigenfunctions already found. In addition, one verifies that
\beq
\Delta \d v_{p,q} = -\mu^2_{p,q}\, \d v_{p,q} \ ,\quad \Delta \d u_{k,l,n} = - M^2_{k,l,n}\, \d u_{k,l,n} \ , \label{Exactoneforms}
\eeq
from which we deduce a basis of exact one-eigenforms. These do not admit a zero-mode, since one should exclude $\d v_{0,0} =0$. Using the orthonormality of the eigenfunctions and proceeding as in (B.5) of \cite{Andriot:2016rdd}, one verifies that these exact one-forms are orthonormal, up to a rescaling by $\mu_{p,q}$ or $M_{k,l,n}$.

We now study the other two pieces of \eqref{hodge}, which are co-closed: this amounts to imposing the condition
\eq{\label{gf}\d^{\dagger}B=0\ \Leftrightarrow \ \d * B =0~.}
If one considers a theory with a gauge symmetry, \eqref{gf} can also be viewed as a gauge-fixing condition. Indeed, for such a theory, the exact piece of $B$ could be removed by a gauge transformation, without loss of generality.

\subsubsection{Co-closed one-forms}\label{sec:oneformsimple}

We expand $B$ as
\eq{\label{bexp}B=\varphi_a(x)e^a~,}
with some scalars $\varphi_a(x)$, $a=1,2,3$. The exterior differential is also expressed on this basis as
\eq{\d=e^aV_a~,}
where the vectors $V_a$, introduced in Section \ref{sec:geoscalar}, are given by
\beq
V_1=\frac{\partial_1}{r^1},\ V_2=\frac{\partial_2}{r^2}-\f x^1 r^1 \frac{\partial_3}{r^3},\ V_3=\frac{\partial_3}{r^3} \ .
\eeq
Using \eqref{Hodgestar}, the co-closed condition \eqref{gf} takes the form
\eq{\label{gf2}
\delta^{ab} V_a \varphi_b=0~.
}
Moreover using \eqref{gf}, and the fact that $\Delta \varphi= \delta^{ab} V_a V_b\, \varphi$ \cite{Andriot:2016rdd}, we compute
\eq{\spl{\Delta B \label{onelapl} &=\left(
\f V_2\varphi_3 + \f V_3\varphi_2-\Delta\varphi_1
\right)e^1 +\left(
-\f V_1\varphi_3 - \f V_3\varphi_1-\Delta\varphi_2
\right)e^2 \\
& +\left(
\f V_1\varphi_2- \f V_2\varphi_1-\Delta\varphi_3 + \f^2 \varphi_3
\right)e^3
~,}}
where we have used \eqref{gf2} and the commutations \eqref{heis} of the $V_a$. The eigenvalue equation \eqref{equation} then gets decomposed on its various components as
\al{
\f V_2 \varphi_3 + \f V_3 \varphi_2 - \Delta \varphi_1 &= \Upsilon \varphi_1 \label{equation1}\\
- \f V_1 \varphi_3 - \f V_3 \varphi_1 - \Delta \varphi_2 &= \Upsilon \varphi_2 \label{equation2}\\
\f V_1\varphi_2- \f V_2\varphi_1-\Delta\varphi_3 + \f^2 \varphi_3 &= \Upsilon \varphi_3 \label{equation3} \ .
}
We now expand our functions $\varphi_a$ on the basis of Laplacian eigenfunctions previously determined,
\eq{\label{fexp}
\varphi_a=\delta_{ab} \sum_{k,l,n} C^b_{k,l,n}\, u_{k,l,n} + \delta_{ab} \sum_{p,q} D^b_{p,q}\, v_{p,q}~,
}
with $C$ and $D$ constant coefficients. The equations to be solved are linear, and the dependence on $x^3$ in $u_{k,l,n}$ is exponential, so this dependence will not get mixed between the two sums above; this is related to the orthogonality of the eigenfunctions. We then treat these two sums independently.

\subsubsection*{Forms independent of $x^3$}

We start with the expansion of the $\varphi_a$ \eqref{fexp} on the $v_{p,q}$. The fact that they are $x^3$-independent simplifies the action of the $V_a$. The condition \eqref{gf2} and the three equations \eqref{equation1}, \eqref{equation2}, \eqref{equation3}, become
\al{
P D^1_{p,q} + Q D^2_{p,q} & = 0 \\
\i \f Q D^3_{p,q} + (P^2 + Q^2) D^1_{p,q} & = \Upsilon D^1_{p,q} \\
- \i \f P D^3_{p,q} + (P^2 + Q^2) D^2_{p,q} & = \Upsilon D^2_{p,q} \\
\i \f P D^2_{p,q} - \i \f Q D^1_{p,q} + (P^2 + Q^2) D^3_{p,q} + \f^2 D^3_{p,q} & = \Upsilon D^3_{p,q} \ ,
}
where we introduce
\beq
P=2\pi \frac{p}{r^1} \ ,\ Q=2\pi \frac{q}{r^2} \ , \  P^2 + Q^2 = \mu_{p,q}^2\ .
\eeq
All solutions (except the trivial $D^{1,2,3}_{p,q}=0$) lead to the following eigenvalues $\Upsilon=Y_{\pm}^{p,q}$
\eq{
Y_{\pm}^{p,q}= P^2+Q^2 + \frac{\f^2}{2} \pm \sqrt{\left( P^2+Q^2 + \frac{\f^2}{2} \right)^2 - (P^2+Q^2)^2} \quad \geq 0 \ , \label{spectrum1formv}
}
which solve the equation
\eq{
(P^2+Q^2) \f^2= (\Upsilon - (P^2+Q^2)-\f^2)(\Upsilon - (P^2+Q^2)) \ .\label{eqUpsipq}
}
The coefficients $D^a_{p,q}$ are fixed by the previous equations, giving the following eigenforms
\beq \label{Bpq}
\spl{
& {\rm For}\ p^2+q^2\neq 0:\quad B^{p,q}_{\pm} = D_{p,q}\, v_{p,q} \left( Q e^1 - P e^2 + \frac{\Upsilon_{\pm}^{p,q} - (P^2+Q^2)}{\i \f } e^3 \right) \\
& {\rm For}\ p=q= 0:\quad Y^{0,0}_-=0:\ B^{0,0}_{1} = D_{0,0}\, v_{0,0} \, e^1\ ,\ B^{0,0}_{2} = D_{0,0}\, v_{0,0} \, e^2 \\
& \phantom{{\rm For}\ p=q= 0:\quad }\ Y^{0,0}_+=\f^2:\ B^{0,0}_{3} = D_{0,0}\, v_{0,0} \, e^3 \ .
}\eeq
Using the orthonormality of the $v_{p,q}$, one verifies non-trivially that the above forms are orthonormal
\beq
\int B^{p,q}_{\epsilon} \w *\, \overline{B^{p',q'}_{\epsilon'}} = \delta_{p,p'} \delta_{q,q'} \delta_{\epsilon,\epsilon'} \ , \label{orthoBpq}
\eeq
upon fixing the normalisation constant $D_{p,q}$, using \eqref{eqUpsipq}, to the value
\beq
D_{p,q} = \frac{1}{\sqrt{\Upsilon_{\epsilon}^{p,q} + P^2+Q^2}}\ {\rm for} \ p\ {\rm or}\ q \neq 0 \ ,\quad D_{0,0}=1 \ . \label{Dpq}
\eeq

\subsubsection*{Forms dependent on $x^3$}

We turn to the expansion of the $\varphi_a$ \eqref{fexp} on the $u_{k,l,n}$, which depend on $x^3$. We first compute from \eqref{uv}, with $y=|\lambda|^{\frac{1}{2}} w_m$, $\lambda= k \frac{2\pi \f}{r^3}$, and the normalisation factor ${\rm norm}_n \propto (2^n n!)^{-\frac{1}{2}}$,
\al{
& \hspace{-0.1in} V_1 u_{k,l,n}= {\rm norm}_n |\lambda|^{\frac{1}{2}} e^{2\pi k \i(x^3+N x^1x^2)} e^{2\pi l \i x^1}\sum_{m\in\mathbb{Z}} e^{2\pi k m\i x^1} |\lambda|^{\frac{1}{4}} e^{-\frac{y^2}{2}}  \i~ {\rm sgn}(\lambda) y H_n(y) \ ,\\
& \hspace{-0.1in} V_2 u_{k,l,n}= {\rm norm}_n |\lambda|^{\frac{1}{2}} e^{2\pi k \i(x^3+N x^1x^2)} e^{2\pi l \i x^1}\sum_{m\in\mathbb{Z}} e^{2\pi k m\i x^1} |\lambda|^{\frac{1}{4}} e^{-\frac{y^2}{2}}  \left(- y H_n(y) + H_n'(y) \right) \ .
}
We use the following properties of the Hermite polynomials $\forall n \in \mathbb{N}$, with $H_{-1}=0$,
\eq{\spl{H_n'(w)&=2n H_{n-1}(w)\\
2w H_n(w)&=H_{n+1}(w)+2n H_{n-1}(w)
~,}\label{hermiteprop}}
to reconstruct the various $u_{k,l,n}$. We get $\forall n \in \mathbb{N}$
\al{
& V_1 u_{k,l,n}= |\lambda|^{\frac{1}{2}} \frac{1}{2} \i~ {\rm sgn}(\lambda) \left( \sqrt{2(n+1)}\, u_{k,l,n+1} + \sqrt{2n}\, u_{k,l,n-1} \right) \ , \label{V1u}\\
& V_2 u_{k,l,n}= |\lambda|^{\frac{1}{2}} \frac{1}{2} \left( - \sqrt{2(n+1)}\, u_{k,l,n+1} + \sqrt{2n}\, u_{k,l,n-1} \right) \ , \label{V2u}\\
& V_3 u_{k,l,n}= \frac{|\lambda|}{\f} \i~ {\rm sgn}(\lambda) u_{k,l,n} \ , \label{V3u}
}
with $u_{k,l,-1}=0$. For convenience, we now change notations with respect to the constant $C$ of \eqref{fexp}, after which we have
\eq{\label{coefc}
\varphi_1= \sum_{n \in \mathbb{N}} c^1_n u_{k,l,n} \sqrt{2^n n!} \ ,\quad \varphi_2= \sum_{n \in \mathbb{N}} \frac{c^2_n}{\i~ {\rm sgn}(\lambda)} u_{k,l,n} \sqrt{2^n n!}\ ,\quad \varphi_3= \sum_{n \in \mathbb{N}} \frac{c^3_n}{2} u_{k,l,n} \sqrt{2^n n!} \ ,
}
where in the new constants $c_n^a$ we drop for simplicity the indices $k,l$ although they should be understood as present.

This material allows us to reformulate the various constraints. We start with the condition \eqref{gf2} that becomes
\eq{
\sum_{n \in \mathbb{N}} |\lambda|^{\frac{1}{2}} \frac{1}{2} \i~ {\rm sgn}(\lambda) \left( c^1_{n-1} + c^2_{n-1}  + 2 (n+1) ( c^1_{n+1} - c^2_{n+1}) + c^3_n \frac{|\lambda|^{\frac{1}{2}}}{\f} \right) u_{k,l,n} \sqrt{2^n n!} = 0 \ ,
}
where we introduced $c^{1,2}_{-1}=0$. Each term of the sum should vanish, leading to, $\forall n \in \mathbb{N}$,
\eq{\label{c3}
c^3_n \frac{|\lambda|^{\frac{1}{2}}}{\f} = - (c^1_{n-1} + c^2_{n-1})  + 2 (n+1) (- c^1_{n+1}  + c^2_{n+1}) \ .
}
We introduce for future convenience $c^{1,2}_{-2}=0$, giving with \eqref{c3} $c^3_{-1}=0$. We turn to the Laplacian equations: \eqref{equation1} and \eqref{equation2} lead respectively to, $\forall n \in \mathbb{N}$
\eq{\spl{
& c^1_n (M^2_{k,l,n} - \Upsilon) + c^2_n |\lambda| + \frac{|\lambda|^{\frac{1}{2}} \f}{4} (2 (n+1) c^3_{n+1} - c^3_{n-1}) = 0 \\
& c^2_n (M^2_{k,l,n} - \Upsilon) + c^1_n |\lambda| + \frac{|\lambda|^{\frac{1}{2}} \f}{4} (2 (n+1) c^3_{n+1} + c^3_{n-1}) = 0 \ ,
}}
where we used \eqref{Lapeq} for the mass. We add and subtract the above two equations, use \eqref{c3}, and obtain $\forall n \in \mathbb{N}$
\al{
(M^2_{k,l,n} - \Upsilon + |\lambda| - (n+1)\f^2)  (c^1_n + c^2_n) - 2 (n+1) (n+2)\f^2 ( c^1_{n+2} - c^2_{n+2}) & = 0 \label{equation1bis}\\
(M^2_{k,l,n} - \Upsilon - |\lambda| + n\f^2)  (-c^1_n + c^2_n) - \frac{\f^2}{2} ( c^1_{n-2} + c^2_{n-2}) & = 0 \ . \label{equation2bis}
}
Finally, using again \eqref{c3}, \eqref{equation3} becomes $\forall n \in \mathbb{N}$
\eq{
(M^2_{k,l,n}  - \Upsilon - |\lambda| + \f^2 )  (c^1_{n-1} + c^2_{n-1}) + 2 (n+1) (M^2_{k,l,n}  - \Upsilon + |\lambda| + \f^2 )  (c^1_{n+1} - c^2_{n+1}) = 0  \ . \label{equation3bis}
}
Introducing
\eq{
\forall n \geq -2,\  c^{\pm}_n=c^1_n \pm c^2_n \ , \label{coefc2}
}
and $c^{1,2}_{-2}=0$, $c^{1,2,3}_{-1}=0$, we summarize the conditions to be solved, \eqref{c3}, \eqref{equation1bis}, \eqref{equation2bis} and \eqref{equation3bis}, as follows $\forall n \in \mathbb{N}$
\al{
& c^3_n \frac{|\lambda|^{\frac{1}{2}}}{\f} = - c^+_{n-1} - 2 (n+1) c^-_{n+1} \label{c3bis}\\
&  c^+_n (M^2_{k,l,n} - \Upsilon + |\lambda| - (n+1)\f^2)  - 2 (n+1) (n+2) \f^2 c^-_{n+2}= 0 \label{equation1ter}\\
&  c^-_{n} (M^2_{k,l,n} - \Upsilon - |\lambda| + n \f^2)  + \frac{\f^2}{2} c^+_{n-2} = 0 \label{equation2ter}\\
&  c^+_{n-1} (M^2_{k,l,n} - \Upsilon - |\lambda| + \f^2)  + 2 (n+1) c^-_{n+1} (M^2_{k,l,n}  - \Upsilon + |\lambda| +\f^2 ) = 0  \ . \label{equation3ter}
}

Equation (\ref{c3bis}) determines the coefficients $c^3_n$ in terms of $c^{1,2}_n$. In their turn the $c^{1,2}_n$ are determined by the system of equations \eqref{equation1ter}-\eqref{equation3ter}, which is overdetermined. To see this more clearly it is convenient to perform a shift in the index $n$, to bring the system to the following form,\footnote{We are looking for a solution to the eigenvalue problem \eqref{equation} with a given $\Upsilon$, and the latter should therefore be considered as fixed. The solution is given by a set of coefficients $c^{1,2,3}_n$. Equations \eqref{equation1ter}-\eqref{equation3ter} involving these coefficients have been obtained by projecting a sum over $n$ on each $u_{k,l,n}$, but the projection could have been done equivalently on each $u_{k,l,n\pm1}$, leading to shifted equations. Doing so would have given equations describing the same solution with eigenvalue $\Upsilon$. So when shifting equations \eqref{equation1ter}-\eqref{equation3ter} as done to reach \eqref{equation1terb}-\eqref{equation3terb}, $\Upsilon$ is considered as fixed.} $\forall n \in \mathbb{N}$,
\al{
&  c^+_n (\alpha_n - (n+1)\f^2)  - 2 (n+1) (n+2) \f^2 c^-_{n+2}= 0 \label{equation1terb}\\
&  c^+_{n}  \f^2+ 2 c^-_{n+2} (\alpha_n+ 2|\lambda| + (n+2) \f^2)  = 0 \label{equation2terb}\\
&  c^+_{n} (\alpha_n + \f^2)  + 2 (n+2) c^-_{n+2} (\alpha_n+ 2|\lambda| +\f^2 ) = 0  \label{equation3terb}
~,}
where we have introduced $\alpha_n=M^2_{k,l,n} - \Upsilon + |\lambda|$, and we have taken into account that $M^2_{k,l,n+p}=M^2_{k,l,n}+2p|\lambda|$.

Equations \eqref{equation1terb}-\eqref{equation3terb} constitute a homogeneous system of three equations for two unknowns. Generically this system can only admit the trivial solution where both $c^+_{n}$ and $c^-_{n+2}$ vanish identically. The necessary and sufficient condition for the existence of a non-trivial solution is the vanishing of all $2\times2$ sub-determinants of the $3\times2$ matrix of coefficients. Remarkably, the three conditions thus obtained turn out to be identical: the system admits non-trivial solutions for $c^+_{n}$, $c^-_{n+2}$, provided $\alpha_n$ obeys the following condition
\eq{\alpha_n^2+\alpha_n(\f^2+2|\lambda|)-2(n+1)|\lambda|\f^2=0~, \label{eqalpha}
}
which ensures that all three equations \eqref{equation1terb}-\eqref{equation3terb} become equivalent. $c^{+}_n$ and $c^{-}_{n+2}$ can then be non-zero, and one is given in terms of the other. From \eqref{eqalpha}, $\alpha_n$ and therefore $\Upsilon$ is determined in terms of $n$; having other coefficients $c^{\pm}_m$, $m\in\mathbb{N}$, would then lead to different $\Upsilon$ and thus correspond to different eigenmodes. For a given $n$, the system is then solved by setting $c^{\pm}_m=0$ for all $m$, except for $c^{+}_n$ and $c^{-}_{n+2}$. This implies that the non-vanishing coefficients of the eigenmode are $c^{1,2}_n$, $c^{1,2}_{n+2}$ and $c^3_{n+1}$. Moreover, those are determined up to an overall constant, corresponding to the normalization of the one-form.

Explicitly, the eigenforms $B_{\pm}^{k,l,n}$ and their eigenvalues $\Upsilon=Y_{\pm}^{k,l,n}$ (we recall $k\in\mathbb{Z}^*$, $n\in\mathbb{N}$, $l=0,\dots, |k|-1$) are given by
\beq \label{Bkln}
\spl{
B_{\pm}^{k,l,n} &=\sum_{a=1}^3 \varphi^{k,l,n}_a\, e^a \\
{\rm where}\ \ \varphi^{k,l,n}_1&=    \frac{1}{2}c\sqrt{2^n n!}\left(
u_{k,l,n}-\frac{\sqrt{(n+1)(n+2)}\f^2}{\alpha_n+2|\lambda|+(n+2)\f^2}~\!u_{k,l,n+2}
\right)
\\
\varphi^{k,l,n}_2&=   -\frac{\i}{2}\text{sgn}(\lambda)c\sqrt{2^n n!}\left(
u_{k,l,n}+\frac{\sqrt{(n+1)(n+2)}\f^2}{\alpha_n+2|\lambda|+(n+2)\f^2}~\!u_{k,l,n+2}
\right)
\\
\varphi^{k,l,n}_3&=
- \frac{1}{2|\lambda|^{\frac{1}{2}}}c\sqrt{2^n n!} ~
\frac{\f\sqrt{2(n+1)}(\alpha_n+2|\lambda|)}{\alpha_n+2|\lambda|+(n+2)\f^2}~\!u_{k,l,n+1}
~,}
\eeq
with $c$ a constant to be determined by the overall normalization, and
\eq{\label{spectrum1formu}
Y_{\pm}^{k,l,n}=M^2_{k,l,n} + 2 |\lambda|
+\frac{1}{2} \f^2
\pm
\sqrt{
\left(|\lambda| +\frac12 \f^2\right)^2
+2(n+1) |\lambda| \f^2
} \quad > 0
~,}
with $ M_{k,l,n}^2 = \tfrac{\lambda^2}{\f^2} + (2n+1) |\lambda|$, or more explicitly
\eq{
Y_{\pm}^{k,l,n}=k^2 \, \left(\frac{2\pi }{r^3}\right)^2 +(2n+3) \frac{2\pi }{r^3} |k \f|
+\frac{1}{2} \f^2
\pm
\sqrt{
\left(\frac{2\pi }{r^3} |k \f| +\frac12 \f^2\right)^2
+2(n+1) \frac{2\pi }{r^3} |k \f| \f^2
}
~.}
As the coefficients $\varphi^{k,l,n}_a$ depend on $\alpha_n=M^2_{k,l,n} - Y_{\pm}^{k,l,n} + |\lambda|$, each of $Y_{\pm}$ leads to a different eigenmode $B_{\pm}$. The orthonormality of the eigenforms is expressed as
\beq
\int B_{\epsilon}^{k,l,n} \w *\, \overline{B_{\epsilon'}^{k',l',n'}} = \int \d^3 x \sqrt{g}\ \delta^{ab}\varphi^{k,l,n,\epsilon}_a \, \overline{\varphi^{k',l',n',\epsilon'}_b}  = \delta_{k,k'} \delta_{l,l'} \delta_{n,n'} \delta_{\epsilon,\epsilon'} \ ,
\eeq
for $\epsilon=\pm$, by appropriately choosing the constant $c$. The orthogonality can be verified using the orthonormality of the $u_{k,l,n}$.\footnote{More generally, one can consider two co-closed one-eigenforms $B_1, B_2$ of eigenvalues $Y_1, Y_2$ in three dimensions. Using that $A\w *B = B\w *A$ for forms of same degree, and integration by parts, one can show
\beq
Y_2\int_3 B_1 \w * \overline{B_2} = \int_3 *B_1 \w \Delta \overline{B_2}= Y_1\int_3 B_1 \w * \overline{B_2}\ .
\eeq
This implies that $B_1$ is orthogonal to $B_2$ if $Y_1\neq Y_2$.}

\subsection{Higher forms and summary}\label{sec:higherforms}

In a three-dimensional space, the spectrum of the two- and three-forms can be deduced respectively from that of the one-forms and the scalars. Indeed, one can always rewrite a $p$-form $A_p$ in terms of its Hodge dual as $A_p= * B_{3-p}$, and one verifies that
\beq
\Delta A_2 = \Upsilon A_2 \Leftrightarrow \Delta B_1 = \Upsilon B_1 \ , \ \Delta A_3 = \Upsilon A_3 \Leftrightarrow \Delta B_0 = \Upsilon B_0 \ .
\eeq
The complete spectrum of scalars and one-forms, as well as the basis of eigenmodes, thus provides those of the two- and three-forms by a simple application of the Hodge star. We summarize the former in Table \ref{tab:summary}.\footnote{Our convention for the Laplacian operator is such that $\Delta = *\, \d*\d + \d*\d\, *$ on any $p$-form, $p\geq 0$. This is the reason for the different signs of the eigenvalues summarized in Table \ref{tab:summary}, where we recall that $Y_{\pm}^{p,q} \geq 0$ and $Y_{\pm}^{k,l,n} > 0$. The more conventional definition, $\Delta = \d^{\dagger}\d + \d \d^{\dagger}$, would provide a positive sign to all eigenvalues.}

\begin{table}[h]
\begin{center}
\begin{tabular}{|c|c|c|}
   \hline
   \multicolumn{2}{|c|}{Eigenmodes}  & Eigenvalues \\
    \hline
    Scalars & $v_{p,q}$ in \eqref{uv} & $-\mu_{p,q}^2$ in \eqref{masses} \\
    \cline{2-3} & $u_{k,l,n}$ in \eqref{uv} & $-M^2_{k,l,n}$ in \eqref{masses} \\
    \hline
    Exact one-forms & $\d v_{p,q}$ with $pq\neq0$ & $-\mu_{p,q}^2$  \\
    \cline{2-3} & $\d u_{k,l,n}$ & $-M^2_{k,l,n}$  \\
    \hline
    Co-closed one-forms & $B_{\epsilon}^{p,q}$ in \eqref{Bpq} & $Y_{\pm}^{p,q}$ in \eqref{spectrum1formv}  \\
    \cline{2-3} & $B_{\epsilon}^{k,l,n}$ in \eqref{Bkln} & $Y_{\pm}^{k,l,n}$ in \eqref{spectrum1formu}  \\
    \hline
\end{tabular} \caption{Scalar and one-form eigenmodes with respective eigenvalues for the Laplacian on the three-dimensional Heisenberg nilmanifold.}\label{tab:summary}
\end{center}
\end{table}

Finally, note we can introduce the following real eigenforms
\beq
B_{r\, \epsilon}^{k,l,n} = \frac{1}{\sqrt{2}} (B_{\epsilon}^{k,l,n} + \overline{B_{\epsilon}^{k,l,n}}) \ , \ B_{r\, \epsilon}^{p,q} = \frac{1}{\sqrt{2}} (B_{\epsilon}^{p,q} + \overline{B_{\epsilon}^{p,q}}) \ . \label{realforms}
\eeq
One has
\beq
\int_3 \d^3 x \sqrt{g}\ u_{k,l,n} u_{k',l',n'} = \delta_{k,-k'} \times \dots  \ ,\ \int_3 \d^3 x \sqrt{g}\ v_{p,q} v_{p',q'} = \delta_{p,-p'} \delta_{q,-q'} \ .
\eeq
This implies that $\{ B_{r\, \epsilon}^{k,l,n}\}$ or  $\{B_{r\, \epsilon}^{p,q} \}$ form an orthonormal set if one restricts e.g.~to $k k'>0$ or $pp' >0, qq'>0$.

\section{Truncation and dimensional reduction on manifolds with SU(3) structure}\label{sec:truncation}

We focus here on reductions of ten-dimensional type II supergravities on manifolds with an SU(3) structure, following \cite{Gurrieri:2002wz, DAuria:2004kwe, House:2005yc, Grana:2005ny, Louis:2006kb, KashaniPoor:2006si}. As will be discussed in more detail in Section \ref{sec:effth}, the starting point is to select a finite set of modes, e.g.~forms on the internal six-dimensional manifold $\mmm$, on which the fields are expanded.  One would ideally like to justify this truncation of the ten-dimensional degrees of freedom to a finite set of modes as leading to an effective action describing the low energy physics of the theory. We do not expect this be to the case here. Rather, we select a finite set of internal forms following a list of conditions that have been identified, especially in \cite{Grana:2005ny, KashaniPoor:2006si}, so that one ends up after dimensional reduction with a four-dimensional ${\cal N}=2$ gauged supergravity. Having an explicit realisation of this program is still interesting and non-trivial, because it is done on manifolds with an SU(3) structure, appearing in type II supergravity backgrounds that are more general than those where $\mmm$ is a Calabi-Yau. The only explicit example where all conditions of \cite{KashaniPoor:2006si} have been satisfied is \cite{KashaniPoor:2007tr}; here, we can make use of our explicit basis of eigenforms on a non-trivial manifold to provide a new concrete example. The reduction made in \cite{Grana:2005ny} mimics to some extent the reduction on Calabi-Yau manifolds, the main difference being that some forms of the finite set are not closed, resulting in non-vanishing SU(3) torsion classes. However, the finite set is closed under the action of the exterior derivative, making the reduction proposed on manifolds with SU(3) structure likely to be a CT, even though there is no general proof of this point.

We follow here the list of conditions of \cite{KashaniPoor:2006si} that completes earlier works, and would like to build a finite set of forms satisfying them. We make use of the Laplacian eigenforms on the three-dimensional nilmanifold, summarized in Section \ref{sec:higherforms}. We consider the six-dimensional $\mmm$ to be the direct product of two three-dimensional compact manifolds, $\mmm = M \times M'$, each of them being either the three-dimensional nilmanifold studied previously or a three-torus.

\subsection{Warm-up: co-closed two-forms}

As part of the requirements on the finite set of internal forms, we need to identify co-closed two-forms $\omega$ on $\mmm$. We start with the case where $\omega$ has two legs on say $M$. It can then be decomposed on the basis of two-eigenforms of the Laplacian on $M$. As discussed in Section \ref{sec:higherforms}, those are the Hodge duals of one-eigenforms. Using the Hodge decomposition, these one-forms are either exact $\d f_I$ or co-closed $B_I$ one-forms, with a general index $I$. Decomposing the two-form on that basis $*_3\d f_I, *_3 B_I$, the coefficients $c_I^e, c_I^c$ are a priori functions on $M'$. Imposing $\omega$ to be co-closed is then equivalent to having $B_I$ harmonic; we rewrite it as $B_I^h$ with coefficient $c_I^h$. The second possibility is that $\omega$ has one leg on each three-dimensional manifold. We then write it as a sum with constant coefficients on wedge products of one-forms, the latter being expanded on the basis of exact $\d f_I$ and co-closed $B_I$. Having $\omega$ co-closed amounts to keeping only the $B_I \w B_J'$.

To summarize, the most general co-closed two-form $\omega$ on $\mmm$ is
\beq
\omega= \sum_I c_I^e *_3\d f_I + c_I^h *_3 B_I^h + c_I^{e'} *_{3'}\d {f_I}' + c_I^{h'} *_{3'}B_I^{h'} + c_{IJ}^{cc'} B_I \w {B_J}'\ ,
\eeq
where $c_I^e, c_I^h$ are functions on $M'$, $c_I^{e'}, c_I^{h'}$ are functions on $M$ and $c_{IJ}^{cc'}$ are constants. We are now interested in its exterior derivative. We first compute
\beq
\d \omega= \sum_I - m^2_{f_I} c_I^e f_I {\rm vol}_3 - m^2_{{f_I}'} c_I^{e'} {f_I}' {\rm vol}_{3'} + c_{IJ}^{cc'} ( \d B_I \w {B_J}' - B_I \w \d {B_J}' ) \ ,\label{d2f}
\eeq
with $\Delta f_I= - m^2_{f_I} f_I$, $*_3 1 = {\rm vol}_3 = \d^3 x \sqrt{g} = e^1\w e^2 \w e^3$, where we set for simplicity all coefficients to be constant. We now turn to the exterior derivative of the co-closed one-forms.

We consider the co-closed one-eigenforms on the nilmanifold, described in Section \ref{sec:oneformsimple}. We first focus on the $B_{\pm}^{k,l,n}$ \eqref{Bkln}, and obtain using \eqref{equation3}
\beq
\d B_{\pm}^{k,l,n} = \tfrac{Y_{\pm}^{k,l,n} - M_{k,l,n+1}^2}{\f}\, \varphi^{k,l,n}_3 \, e^1\w e^2 + (-V_3 \varphi^{k,l,n}_1 + V_1 \varphi^{k,l,n}_3 )\, e^1\w e^3 + (-V_3 \varphi^{k,l,n}_2 + V_2 \varphi^{k,l,n}_3 )\, e^2\w e^3 \ . \nn
\eeq
Properties \eqref{V1u} - \eqref{V3u} are used to compute the remaining terms. In addition, we verify using \eqref{eqalpha} the identity
\beq
\frac{\f(\alpha_n+2|\lambda|) (n+1)}{\alpha_n+2|\lambda|+(n+2)\f^2} =  \frac{\alpha_n}{\f} \ .
\eeq
This allows to show that
\beq
\d B_{\pm}^{k,l,n} = \frac{Y_{\pm}^{k,l,n} - M_{k,l,n}^2 - 2 |\lambda|}{\f} *_3 B_{\pm}^{k,l,n} \ .
\eeq
The coefficient is equal $-(\alpha_n + |\lambda|)/\f$. Another identity verified by virtue of \eqref{eqalpha} is
\beq
\frac{(\alpha_n + |\lambda|)^2}{\f^2} = Y_{\pm}^{k,l,n} \ .
\eeq
The above is then rewritten as
\beq
\d B_{\pm}^{k,l,n} = - \sqrt{Y_{\pm}^{k,l,n}}\ *_3 B_{\pm}^{k,l,n} \ .\label{d1f1}
\eeq
This is clearly compatible with the following identity, valid for any co-closed one-form in three dimensions with $\Delta B_I = \Upsilon B_I$,
\beq
\int_3 \d B_I \w *_3 \d B_I = \Upsilon \int_3 B_I \w *_3 B_I  \ , \label{idBI}
\eeq
shown with an integration by parts. This identity also shows that the Laplacian eigenvalue of a real co-closed one-form in three dimensions has to be positive. Another consistency check is the following result for a constant $a$
\beq
\d B_I = a * B_I \ ,\ \d *B_I = 0\ ,\ \Delta B_I = \Upsilon B_I \ \Rightarrow \ \Upsilon = a^2 \ ,\label{check}
\eeq
shown by applying $* \d *$ on both sides of the first equality.

We turn to the other co-closed one-eigenforms on the nilmanifold $B_{\epsilon}^{p,q}$ \eqref{Bpq}. From \eqref{eqUpsipq}, one deduces the following identity
\beq
(Y_{\pm}^{p,q} - (P^2+Q^2))^2 = \f^2 Y_{\pm}^{p,q} \ .
\eeq
Using this, one shows
\beq
\d B_{\pm}^{p,q} = \pm {\rm sgn}(\f)  \sqrt{Y_{\pm}^{p,q}}\ *_3 B_{\pm}^{p,q} \ .\label{d1f2}
\eeq
One verifies that this formula holds as well for $p=q=0$. This result is again compatible with \eqref{idBI} and \eqref{check}. The properties \eqref{d1f1} and \eqref{d1f2} combined with the exterior derivative of the co-closed two-form \eqref{d2f} will be useful in the following.

\subsection{A first set of forms}

Inspired by the results of the previous section, we now build a set of forms verifying some of the conditions listed in \cite{KashaniPoor:2006si}. We consider co-closed one-eigenforms $\{ B_I \}$ on $M$ and $\{ {B_I}' \}$ on $M'$. We restrict to orthonormal and real one-forms, which in addition verify
\beq
\d B_I = s_I \sqrt{Y_I} *_3 B_I \ ,\label{dB*B}
\eeq
and similarly for ${B_I}'$, where $s_I=\pm 1$ is a sign and $Y_I$ is the real, positive, Laplacian eigenvalue, as in \eqref{d1f1} and \eqref{d1f2}. Such one-forms can be built from the previous examples: see \eqref{realforms} and below. We now introduce the following set of forms
\beq \label{genset}
\spl{
& \omega_{IJ} = \frac{1}{{\cal N}_{IJ}} B_I \w {B_J}' \ , \quad \tilde{\omega}^{KL} = ({\cal N}_{IJ})^2 \delta^{KI} \delta^{LJ} *_6 \omega_{IJ} = {\cal N}_{IJ} \delta^{KI} \delta^{LJ} *_3 B_I \w *_{3'} {B_J}'\ , \\
& \alpha_{IJ} = \frac{1}{{\cal A}_{IJ}} {B_J}' \w *_3 B_I  \ ,\quad \beta^{KL} = - {\cal A}_{IJ}\, \delta^{KI} \delta^{LJ} B_I \w *_{3'} {B_J}' \ ,
}\eeq
where ${\cal N}_{IJ}$ and ${\cal A}_{IJ}$ are real normalization constants, whose indices should not be summed over. The orientation convention, necessary when splitting the $*_6$ on each three-dimensional space, goes as follows: the six indices are ordered as the three of $M$ first, followed by the three of $M'$. This set of forms verifies the following properties
\bea
& \int_6 \omega_{IJ} \w *_6\, \omega_{KL} = \frac{1}{({\cal N}_{IJ})^2} \delta_{IK} \delta_{JL}\\
& \int_6 \alpha_{IJ} \w \beta^{KL} = \delta_I^K \delta_J^L \ , \ \int_6 \alpha_{IJ} \w \alpha_{KL} = \int_6 \beta^{IJ} \w \beta^{KL} = 0 \\
& *_6 \alpha_{IJ} = B_{IJ,KL} \beta^{KL} \ , \  *_6  \beta^{KL} = C^{KL,MN} \alpha_{MN} \\
& {\rm where}\ B_{IJ,KL} = ({\cal A}_{KL})^{-2} \delta_{IK} \delta_{JL} \ ,\ C^{KL,MN} = - ({\cal A}_{MN})^2 \delta^{KM} \delta^{LN} \\
& *_6 \d *_6 \omega_{IJ} = 0\ ,\ \d \tilde{\omega}^{KL} = 0 \\
& \d \omega_{IJ} = m_{IJ}{}^{KL} \alpha_{KL}  + e_{IJ,KL} \beta^{KL} \ ,\ \d \alpha_{IJ} = e_{KL,IJ}\tilde{\omega}^{KL} \ ,\ \d \beta^{KL} = -  m_{IJ}{}^{KL} \tilde{\omega}^{IJ} \\
& {\rm where}\ m_{IJ}{}^{KL} = \frac{s_I \sqrt{Y_I} {\cal A}_{IJ}}{{\cal N}_{IJ}} \delta^K_I \delta^L_J \ ,\ e_{IJ,KL} = \frac{{s_J}' \sqrt{{Y_J}'} }{{\cal N}_{IJ} {\cal A}_{IJ}} \delta_{IK} \delta_{JL} \\
&  m_{IJ}{}^{KL} e_{MN,KL} - e_{IJ,KL} m_{MN}{}^{KL} = 0 \ .
\eea
In addition, the entries of $m$ and $e$ can be made integer, by choosing for instance
\bea
& {\rm For}\ Y_I{Y_J}' \neq 0:\quad  {\cal A}_{IJ} = \left(\frac{{Y_J}'}{Y_I} \right)^{\frac{1}{4}}\ , \ {\cal N}_{IJ}= \frac{(Y_I{Y_J}')^{\frac{1}{4}}}{N_{IJ}} \ ,\ N_{IJ} \in \mathbb{Z}^* \ ,\label{normaprop}\\
& {\rm For}\ Y_I=Y \neq 0\ ,\ {Y_J}'=0\ \mbox{or vice-versa}:\quad  {\cal A}_{IJ} = 1\ , \ {\cal N}_{IJ}= \frac{\sqrt{Y}}{N_{IJ}} \ ,\ N_{IJ} \in \mathbb{Z}^* \ .
\eea
This way, conditions 1,2,3,5 of \cite{KashaniPoor:2006si} are satisfied.

Only conditions related to the SU(3) structure remain. One has to build the SU(3) structure forms $J$ and $\Omega$ in terms of a finite set of forms among the previous ones: $J$ in terms of the $\omega$ and $\Omega$ in terms of the $\alpha$ and $\beta$. The forms $J$ and $\Omega$ have to satisfy certain constraints, among which the compatibility condition $J\w\Omega=0$. A way to ensure this is to impose, as in \cite{Grana:2005ny},
\beq
\omega \w \alpha = \omega\w \beta = 0 \ , \label{compat}
\eeq
for all $\omega, \alpha, \beta$ entering $J$ and $\Omega$, while a refined constraint is considered in condition 6 of \cite{KashaniPoor:2006si}. For illustration, we consider our forms $\omega_{II}, \alpha_{II}, \beta^{II}$: applying the constraint \eqref{compat} then amounts to finding $B_I, B_J, I\neq J$ such that $B_I \w *_3 B_J=0$. Except in the case to be treated in Section \ref{sec:su3}, this is difficult to achieve with our co-closed one-forms, due to their various components, but also because of the functions appearing and the absence of an integral in that constraint. We still find one solution by taking $B_1^{0,0}, B_{r \, \epsilon}^{p,0}, p\neq 0$ (the latter is defined in \eqref{realforms}) because $B_{r\, \epsilon}^{p,0} \w *_3 B_1^{0,0} = 0$; one can equivalently take $B_2^{0,0}, B_{r\, \epsilon}^{0,q}$, but not all four forms together. From those, one can build two different forms $\omega_{II}$. Another condition of the SU(3) structure is however
\beq
\int_6 J \w J\w J \neq 0 \ .
\eeq
The two $\omega_{II}$ built from $B_1^{0,0}, B_{r\, \epsilon}^{p,0}$ are then not enough. We now turn to a simpler option to build the SU(3) structure forms.

\subsection{Finite set of forms and SU(3) structure}\label{sec:su3}

We consider the following co-closed, real, orthonormal one-eigenforms on $M$
\beq
B_1= v_{0,0}\, e^1 \ ,\ B_2= v_{0,0}\, e^2 \ ,\ B_3= v_{0,0}\, e^3 \ ,
\eeq
and similarly for $M'$ forms. One has
\beq
Y_1=Y_2={Y_1}'={Y_2}'=0 \ ,\ Y_3=\f^2 \ ,\ {Y_3}'={\f'}^2 \ ,\ s_3= {\rm sgn}(\f) \ ,\ {s_3}'= {\rm sgn}(\f') \ .
\eeq
In that case, one treats at the same time the three-torus and the three-dimensional nilmanifold by setting or not the structure constant(s) to zero. Of the forms of \eqref{genset}, we only need those with twice the same index, i.e.~$\omega_{II}, \alpha_{II}$, etc. We then replace the doubled index by only one, for $I=1,2,3$, as follows
\beq
\spl{
& \omega_{I} = \frac{1}{{\cal N}_{I}} B_I \w {B_I}' \ , \quad \tilde{\omega}^{I} = {\cal N}_{I} *_3 B_I \w *_{3'} {B_I}'\ , \\
& \alpha_{I} = \frac{1}{{\cal A}_{I}} {B_I}' \w *_3 B_I  \ ,\quad \beta^{I} = - {\cal A}_{I}\, B_I \w *_{3'} {B_I}' \ ,
}\eeq
where the expressions are understood without sum over indices. We introduce in addition the following forms to complete our finite set, with some real constant ${\cal A}_{0}$
\beq
\spl{
& \alpha_0 = - \frac{{ v_{0,0}}'}{ v_{0,0}} \frac{1}{{\cal A}_{0}}\, {B_1}' \w *_3 {B_1}' = - \frac{1}{ v_{0,0} {\cal A}_{0}}\, {B_1}' \w {B_2}' \w {B_3}' \ ,\\
& \beta^0 = \frac{ v_{0,0}}{{ v_{0,0}}'} {\cal A}_{0}\, B_1 \w *_3 B_1 = \frac{{\cal A}_{0}}{{ v_{0,0}}'}\,  B_1 \w B_2 \w B_3 \ .
}\eeq
These new forms verify all previous requirements extended to a new index 0, namely for $I=1,2,3$,
\bea
& \int_6 \alpha_{0} \w \beta^{0} = 1 \ , \ \int_6 \alpha_{0} \w \alpha_{0} = \int_6 \beta^{0} \w \beta^{0} = 0 \\
& \int_6 \alpha_{0} \w \beta^{KL} = \int_6 \alpha_{IJ} \w \beta^{0} = 0 \ , \ \int_6 \alpha_{0} \w \alpha_{KL} = \int_6 \beta^{0} \w \beta^{KL} = 0 \nn \\
& *_6 \alpha_{0} = B_{0,0} \beta^{0} \ , \  *_6  \beta^{0} = C^{0,0} \alpha_{0}\ {\rm where}\ B_{0,0} = \left( \frac{{ v_{0,0}}'}{v_{0,0}} \right)^4 ({\cal A}_{0})^{-2} \ ,\ C^{0,0} = - \left( \frac{v_{0,0}}{{ v_{0,0}}'} \right)^4 ({\cal A}_{0})^2  \nn\\
& B_{IJ,0} = B_{0,KL} = 0 \ ,\ C^{0,MN} = C^{KL,0} = 0 \nn\\
& m_{IJ}{}^{0} = e_{IJ,0} =0 \ ,\ \d \alpha_{0} = \d \beta^{0} = 0 \ ,\nn
\eea
where we have introduced new coefficients for $B, C, m, e$. A generalization of these new forms, ${B_I}' \w *_3 {B_I}'$ and $B_I \w *_3 B_I$ , could have been introduced previously with the more general set \eqref{genset}.

We now define the SU(3) structure forms
\beq
Z_I= B_I + \i {B_I}' \ ,\quad J=\frac{\i}{2} \sum_{I=1}^{3} Z_I \w \overline{Z_I} \ ,\quad \Omega= Z_1 \w Z_2 \w Z_3 \ .
\eeq
Using the above, one shows
\beq
\spl{
& J= \sum_{I=1}^{3} V^I \omega_I \ ,\quad \Omega= \sum_{I=0}^{3} X^I \alpha_I - G_I \beta^I \ ,\\
& {\rm where}\ V^I= {\cal N}_{I} , \ X^I=\i v_{0,0} {\cal A}_{I} \ ,\ G^I= -\frac{{ v_{0,0}}' }{ {\cal A}_{I}} \ .
}\eeq
With our forms, one verifies \eqref{compat}, i.e.
\beq
\forall I=1,2,3, J=0,1,2,3\ ,\ \omega_I \w \alpha_J = \omega_I \w \beta^J = 0 \ .
\eeq
Condition 6 of \cite{KashaniPoor:2006si} is then trivially satisfied, and all conditions of \cite{Grana:2005ny} are verified. We are left with additional requirements found in \cite{KashaniPoor:2006si}, namely conditions 4,7,8,9, that will be verified by our finite set of forms and the above SU(3) structure.

The remaining conditions have to do with moduli dependence. Satisfying them requires to fix the normalization constants. We choose
\beq
\spl{
& {\cal N}_1=\sqrt{\frac{r^1 {r^1}'}{r^2{r^2}'r^3{r^3}'}}\frac{1}{N_1} \ ,\ {\cal N}_2=\sqrt{\frac{r^2 {r^2}'}{r^1{r^1}'r^3{r^3}'}}\frac{1}{N_2} \ ,\ {\cal N}_3=\sqrt{\frac{r^3 {r^3}'}{r^1{r^1}'r^2{r^2}'}}\frac{1}{N_3} \ ,\ N_I \in \mathbb{Z}^* \ ,\\
& {\cal A}_1= \sqrt{\frac{r^2 r^3 {r^1}'}{{r^2}'{r^3}' r^1}} \ ,\ {\cal A}_2= \sqrt{\frac{r^1 r^3 {r^2}'}{{r^1}'{r^3}' r^2}}\ ,\ {\cal A}_3= \sqrt{\frac{r^1 r^2 {r^3}'}{{r^1}'{r^2}' r^3}}  \ ,\ {\cal A}_0= \sqrt{\frac{r^1 r^2 r^3}{{r^1}'{r^2}'{r^3}'}} = \frac{{v_{0,0}}'}{v_{0,0}} \ ,
}\eeq
where we recall that
\beq
v_{0,0}=\frac{1}{\sqrt{V}}=\frac{1}{\sqrt{r^1r^2r^3}} \ , \ {v_{0,0}}'=\frac{1}{\sqrt{V'}}=\frac{1}{\sqrt{{r^1}'{r^2}'{r^3}'}} \ .
\eeq
This is motivated by the requirement of having integer $m$ and $e$ coefficients,\footnote{The choice made corresponds to the suggestion \eqref{normaprop}, except for the integers $N$ and $N'$ of the structure constants that are not included in the normalization constants.} given here as follows for $K=0,1,2,3$
\beq
m_1{}^K=m_2{}^K= e_{1K}=e_{2K}= 0 \ , \ m_3{}^K= \delta_3^K\, N\, N_3 \ ,\ e_{3K}=\delta_{3K}\, N' \, N_3 \ .
\eeq
A further motivation comes from condition 8 of \cite{KashaniPoor:2006si}, stating that the following integral should not depend on moduli, e.g.~here the radii
\beq
\int_6 \omega_I \w \omega_J \w \omega_K = - 6\, \delta_{(I}^1 \delta_J^2 \delta_{K)}^3\,  N_1 N_2 N_3 \ ,
\eeq
the condition 8 being therefore satisfied here. In addition, these normalization constants make our set of forms, $\omega_{I=1,2,3},\ \tilde{\omega}^{I=1,2,3}$ and $\alpha_{K=0,1,2,3},\ \beta^{K=0,1,2,3}$, completely independent of any radii, when expressed in terms of the $\d x^m$. Conditions 7 and 9 of \cite{KashaniPoor:2006si} are then also satisfied.

We are left with condition 4 of \cite{KashaniPoor:2006si}: it requires, in our framework, the forms
\beq
\phi_I= \alpha_I - \del_I G_J \beta^J \ ,\ I=0,1,2,3
\eeq
to be (3,0) and (2,1) only, where we interpret this in terms of the almost complex structure defined by the $\{Z_I,\overline{Z_I}\}$. The coefficients $\del_I G_J$ are defined thanks to $\del_I=\frac{\del}{\del X^I}$, a derivative we elaborate on below. To determine the type of forms we have, we rewrite the $B_I$ and ${B_I}'$ in terms of the $Z_I$ and $\overline{Z_I}$. For instance, one gets
\beq
\spl{
\alpha_1 = \frac{1}{{\cal A}_1 8 \i v_{0,0}} &\big( Z_{123} - \overline{Z_{123}} + Z_{12}\w \overline{Z_3} - Z_{13}\w \overline{Z_2} - Z_{23}\w \overline{Z_1} \\
& + Z_{1}\w \overline{Z_{23}} + Z_{2}\w \overline{Z_{13}} - Z_{3}\w \overline{Z_{12}} \big)\ ,
}\eeq
where we denote $Z_{123}= Z_1 \w Z_2 \w Z_3$, etc. There is a unique combination of $\alpha_1$ and $\beta^I$ that contains no (0,3) and (1,2) terms: it is given by
\beq \label{combi21}
\spl{
& \alpha_1 - \frac{\i}{2} \frac{{v_{0,0}}'}{v_{0,0}  {\cal A}_1} \left( \frac{\beta^0}{{\cal A}^0} - \frac{\beta^1}{{\cal A}^1} + \frac{\beta^2}{{\cal A}^2} + \frac{\beta^3}{{\cal A}^3}  \right) \\
& = \frac{1}{4\i{\cal A}_1 v_{0,0}} \left( Z_{123} + Z_{12}\w \overline{Z_3} -  Z_{13}\w \overline{Z_2} -  Z_{23}\w \overline{Z_1} \right) \ .
}\eeq
The question is now whether the coefficients appearing in the above correspond to $-\del_1 G_J$. Using the relation
\beq
X^0 X^1 X^2 X^3 = (v_{0,0} {v_{0,0}}')^2 \ ,
\eeq
one shows that
\beq
G_I= - \sqrt{-\frac{X^J X^K X^L}{X^I}} \ ,\ {\rm for} \ \epsilon_{IJKL}\neq 0 \ .
\eeq
We deduce the following results
\beq
\del_J G_I = \frac{\i}{2} \frac{{v_{0,0}}'}{v_{0,0}} \frac{1}{{\cal A}_I {\cal A}_J}\ {\rm for}\ J\neq I \ ,\quad  \del_I G_I = -\frac{\i}{2} \frac{{v_{0,0}}'}{v_{0,0}} \frac{1}{{\cal A}_I^2 } \ .
\eeq
This matches precisely the coefficients in the combination \eqref{combi21}. One verifies explicitly that the same holds for the other forms, i.e.~$\phi_{I=0,1,2,3}$ are only (3,0) and (2,1) forms. Condition 4 of \cite{KashaniPoor:2006si} is then satisfied.

To conclude, our finite set of forms $\omega_{I=1,2,3}, \tilde{\omega}^{I=1,2,3}, \alpha_{J=0,1,2,3}, \beta^{J=0,1,2,3}$, together with the above SU(3) structure, verify all conditions of \cite{Gurrieri:2002wz, DAuria:2004kwe, House:2005yc, Grana:2005ny, Louis:2006kb, KashaniPoor:2006si}, so that one eventually obtains from this truncation a four-dimensional ${\cal N}=2$ gauged supergravity. It provides a new explicit example satisfying all conditions of \cite{KashaniPoor:2006si}, the only other known example being that of \cite{KashaniPoor:2007tr} (the work \cite{Cassani:2009ck} presents interesting examples satisfying conditions 1-6 of \cite{KashaniPoor:2006si}). Note that our forms are built from Maurer--Cartan forms with constant coefficients: this truncation is then expected to be consistent and give a gauged supergravity, see Section \ref{sec:effth}. Furthermore, our set of forms (together with the constant function and the six-dimensional volume form) turn out to correspond to the set of even forms under a certain projection, namely that of three space-filling orientifold $O_5$-planes, wrapping respectively the internal directions $e^I \w e^{I'}$ given by $\omega_{I=1,2,3}$. Indeed our forms can be written in terms of parallel or transverse directions as $e^{||}\w e^{||},\ e^{\bot}\w e^{\bot}, \ e^{||}\w e^{\bot}\w e^{\bot}$. We do not know of a reference where this precise reduction has been performed towards a gauged supergravity, so our set of forms provides a new result on such a reduction. This remark on the projection may also indicate a way to go from the finite set of forms obtained by a low energy truncation in Section \ref{sec:light}, which gives a trivial structure group (the manifold is parallelizable), to the present, more restricted finite set (in particular having no one- or five-forms) from which one builds the SU(3) structure. Such a relation would be interesting: it would provide a justification of the present truncation and finite set, as capturing the low energy physics.

\section{Low energy approximation, and the swampland}\label{sec:lowenergy}

\subsection{The light spectrum}\label{sec:light}

The purpose of having fluxes in compactifications from ten to four dimensions is often to stabilize moduli, i.e.~providing them with a mass. One usually needs to truncate the infinite towers of Kaluza--Klein modes, so keeping the flux energy scale requires that it should be below the first Kaluza--Klein mass, if one wants the truncation to make sense as a low energy approximation. Therefore, having a few light massive modes in addition to the massless ones, with a mass scale given by the fluxes, requires a hierarchy between fluxes and Kaluza--Klein masses. In the case of a torus, the hierarchy is given by the large volume limit, which is also the supergravity or classical limit. Indeed, a flux like the $H$-flux is quantized as follows
\beq
\frac{1}{4\pi^2 \alpha'} \int_3 H = \frac{1}{4\pi^2 l_s^2} \int_3 H_{123}\, e^1 \w e^2 \w e^3 =N \in \mathbb{Z}^*
\eeq
which gives, for a constant flux on a three-torus (with radii $R^{m=1,2,3}$)
\beq
H_{123} = N \frac{l_s^2}{2\pi R^1 R^2 R^3}\ .
\eeq
The energy scale of the flux $H_{123}$ can then be made much smaller than the Kaluza--Klein mass $1/R^{m=1,2,3}$ if $R^{m=1,2,3} \gg l_s \, (\times \sqrt{|N|})$, i.e.~if there is a large volume. This hierarchy may provide a justification for the usual truncation made on Calabi-Yau manifolds, even though knowing the precise low energy theory still requires to study fluctuations around an explicit background, as discussed in Section \ref{sec:effth}.

Here we present a different limit or approximation, that generates analogously a hierarchy between the ``geometric flux'' $\f$ and the Kaluza--Klein scales $1/r^m$, therefore allowing to keep the former while truncating the latter. We recall that
\beq
\f = \frac{N r^3}{r^1 r^2} \ ,\ N \in \mathbb{Z}^* \ ,\ r^m >0 \ .
\eeq
We propose to consider the following approximation
\beq
|N| r^3 \ll r^1 \ ,\ |N| r^3 \ll r^2 \label{approx} \ ,
\eeq
which can be understood (it implies $r^3 \ll r^{1,2}$) as having a small fiber (along $e^3$) compared to the base (along $e^{1,2}$); one may refer to this as a small fiber, or large base, limit, in analogy to the large volume limit. This regime can be motivated from the T-dual setup of an $H$-flux on a torus with large volume. Indeed, the T-duality rules give
\beq
H_{123} = - f^3{}_{12} = \f \ ,\ r^1=2\pi R^1\ ,\ r^2=2\pi R^2 \ ,\ r^3=2\pi \frac{l_s^2}{R^3} \ ,
\eeq
where we rescale the new radii by $2\pi$ to fit our conventions; the condition $R^{1,2,3} \gg l_s \sqrt{|N|}$ then becomes $ |N| r^3 \ll 2\pi l_s \sqrt{|N|} \ll r^{1,2}$, from which one recovers \eqref{approx}. The T-dual picture is however only a motivation, as we rather require here $ 2\pi l_s |N| \ll |N| r^3 \ll  r^{1,2}$, to remain in the supergravity (and large volume) regime. It is physically plausible to have a small fiber, for instance in the class of solutions of \cite{Andriot:2016ufg} where branes and orientifolds wrap the fiber.

We deduce from \eqref{approx} the following hierarchies
\beq
|\f| \ll \frac{1}{r^1} , \ \frac{1}{r^2} \ll \frac{1}{r^3} \ . \label{frhierarchy}
\eeq
As anticipated, the geometric flux generates a light energy scale compared to the base Kaluza--Klein scale, itself light compared to the fiber Kaluza--Klein scale. We thus introduce the following low energy approximation or truncation
\beq
\mbox{Low energy approximation:}\quad \mbox{truncate modes of mass $\geq\ \tfrac{1}{r^1}, \tfrac{1}{r^2}, \tfrac{1}{\sqrt{r^1 r^2}}$, given \eqref{approx}} .\label{LEapprox}
\eeq
We now determine the resulting spectrum of light modes, using the summary of Section \ref{sec:higherforms}.

\subsection*{Light scalars and three-forms}

With the previous notation
\beq
P=2\pi \frac{p}{r^1} \ ,\ Q=2\pi \frac{q}{r^2}  \ ,\ \lambda= k \frac{2\pi \f}{r^3} \ ,
\eeq
we rewrite the squared masses of the scalar spectrum \eqref{masses}
\beq
\mu^2_{p,q} = P^2 + Q^2 \ ,\qquad M_{k,l,n}^2 = \frac{\lambda^2}{\f^2} + (2n+1) |\lambda| \ .
\eeq
We recall that $p,q,n \in \mathbb{N}$, $k \in \mathbb{Z}^*$. Therefore, the condition \eqref{frhierarchy} implies
\beq
\f^2 \ll |\lambda| \ll \frac{\lambda^2}{\f^2} \ . \label{flambdahier}
\eeq
Given that $|\lambda|= |kN|2\pi \tfrac{1}{r^1 r^2}$, we deduce that the low energy approximation \eqref{LEapprox} only leaves the scalar mode $v_{0,0}$ of mass $\mu_{0,0}=0$, all others are truncated. Correspondingly, the only light three-form is $v_{0,0} \, e^1\w e^2 \w e^3$.

\subsection*{Light one- and two-forms}

We start with the exact one-forms. Their spectrum is that of the scalar eigenmodes, without the zero-mode: there is therefore no light mode among those forms. We turn to the co-closed one-forms, and first consider $B_{\epsilon}^{p,q}$, whose spectrum is given in \eqref{spectrum1formv}. For $P^2 + Q^2 \neq 0$, one can develop its expression into
\beq
Y^{p,q}_{\pm} = (P^2 + Q^2) \left( 1 \pm \frac{|\f|}{\sqrt{P^2 + Q^2}} + \frac{1}{2} \frac{\f^2}{P^2 + Q^2} + o\left(\frac{\f^2}{P^2 + Q^2} \right)  \right) \ .
\eeq
Therefore, for $P^2 + Q^2 \neq 0$, the low energy approximation \eqref{LEapprox} truncates this whole spectrum. We are left with the following light modes
\beq
\spl{
&B^{0,0}_{1} = v_{0,0}\, e^1, B^{0,0}_{2} = v_{0,0}\, e^2, \ \ Y^{0,0}_-=0 \ ,\\
&B^{0,0}_{3} = v_{0,0}\, e^3,\qquad \qquad \qquad \ \ Y^{0,0}_+= \f^2 \ .
}\eeq
Correspondingly, the light two-forms are $v_{0,0} \, e^2 \w e^3$, $v_{0,0} \, e^3\w e^1$, $v_{0,0} \, e^1\w e^2$.

We turn to the one-forms $B_{\epsilon}^{k,l,n}$, whose spectrum is given in \eqref{spectrum1formu}. We introduce the dimensionless parameter
\beq
\ep= \frac{\f^2}{|\lambda|}
\eeq
which is small compared to $1$ given \eqref{flambdahier}. We first rewrite the eigenvalue as
\eq{
Y_{\pm}^{k,l,n}=|\lambda| \left( \epsilon^{-1} + 2n + 3 + \frac{1}{2} \epsilon \pm \sqrt{ 1 + \epsilon (2n+3) + \frac{1}{4} \epsilon^2 } \right)
~.}
Even though $\epsilon \ll 1$, one should pay attention to quantities like $\epsilon\, n$, as $n$ can be arbitrarily big. However, one has
\beq
\epsilon^{-2} \gg 1 \ , \qquad (2n + 3)^2 \gg \epsilon (2n+3) \ ,
\eeq
so the square root can be neglected, and one eventually gets
\eq{
Y_{\pm}^{k,l,n}=|\lambda| \left( \epsilon^{-1} + 2n + 3 + o(\epsilon^{-1} + 2n + 3) \right)
~.}
Since $|\lambda| \epsilon^{-1} = k^2 \, \left(\frac{2\pi }{r^3}\right)^2 $, we conclude that this whole spectrum is truncated by the low energy approximation \eqref{LEapprox}, and no light state remains.

\subsection*{Summary: the light modes}

To summarize, the low energy approximation \eqref{LEapprox} truncates the spectrum to the following light modes on the nilmanifold, given as set of forms with their eigenvalue in brackets
\eq{\label{lightmodes}
\spl{&v_{0,0}\ (0),\  v_{0,0}\, e^1 \ (0), \ v_{0,0}\, e^2 \ (0),\ v_{0,0}\, e^3 \ (\f^2), \\
&v_{0,0} \, e^2 \w e^3 \ (0),\ v_{0,0} \, e^3\w e^1 \ (0),\ v_{0,0} \, e^1\w e^2 \ (\f^2),\ v_{0,0}\, e^1\w e^2 \w e^3\ (0)\ .}
}
Up to the normalisation constant $v_{0,0} =\tfrac{1}{\sqrt{V}}$, this turns-out to be the complete set of forms built from the Maurer--Cartan one-forms, with constant coefficients.

\subsection{Different regimes: entering the swampland}\label{sec:swamp}

Before discussing in Section \ref{sec:effth} the effective theory associated to the previous low energy truncation, let us explore what happens when going away from this regime \eqref{approx}, that allowed the hierarchy \eqref{frhierarchy} and truncation \eqref{LEapprox}. The three radii $r^m$ play here the role of ``moduli'', and we now move away, in the corresponding moduli space or field space, from this point where we found a controlled truncation to a finite set of light modes. This is motivated by the recent discussion on the validity of effective theories, when one moves at large distances (Planck scale) in field space. For effective theories of a quantum gravity, the following behavior has been conjectured \cite{Ooguri:2006in, Klaewer:2016kiy}: if one moves from a point $\phi_0$ by a field space distance $\Delta \phi$, there will be an (infinite) tower of states of mass $m(\phi)$ that become exponentially light as follows
\beq
m(\phi_0 + \Delta \phi) = m(\phi_0) \ f(\phi_0,\Delta \phi) \ e^{-\alpha \frac{\Delta \phi}{M_p}} \ , \label{conj}
\eeq
where $M_p$ is the Planck mass, $f$  is a subdominant function with respect to the exponential, and $\alpha$ should be of order $1$. If this ``refined swampland distance conjecture'' holds, it implies that the quantum gravity effective theory at $\phi_0$, where the tower of states has been truncated, would not be a valid description anymore after $\Delta \phi \sim M_p$. This would have consequences for e.g.~large field inflation models that allow such displacements; we refer to \cite{Blumenhagen:2018hsh} for a recent review, while various checks and discussions on this conjecture can be found in \cite{Valenzuela:2016yny, Blumenhagen:2017cxt, Palti:2017elp, Hebecker:2017lxm, Cicoli:2018tcq, Grimm:2018ohb, Heidenreich:2018kpg, Blumenhagen:2018nts, Landete:2018kqf}. Testing the proposal \eqref{conj} presents two independent difficulties: first, one needs to know the dependence of the spectrum on the fields or moduli, $m(\phi)$, and second, one needs to know the proper field space distance $\Delta \phi$. The latter requires the field space metric, which can be read off of the kinetic terms. Here, we make use of our knowledge of the Laplacian spectrum to discuss the first point: even though it does not necessarily coincide with the mass spectrum of the theory (see Section \ref{sec:effth}), it still provides a first intuition on the various towers of modes.

Our starting $\phi_0$ is the regime \eqref{approx} where none of $r^{m=1,2,3}$ is large, so that the Kaluza--Klein towers are truncated as in \eqref{LEapprox}. If $r^1$ or $r^2$ becomes large, many modes will become light, starting with the scalars $v_{p,q}$ which admit the standard Kaluza--Klein spectrum. Let us rather maintain $r^1, r^2$ fixed. To reach a different regime, we then send $r^3$ to be (very) large. This amounts to
\beq
r^3 \gg r^1, r^2 \quad \Rightarrow \quad |\f| \gg \frac{1}{r^1}, \frac{1}{r^2} \gg \frac{1}{r^3} \ .
\eeq
Interestingly, $\tfrac{|\f|}{r^3}= \tfrac{|N|}{r^1 r^2}$ remains finite with $r^1, r^2$ fixed (but not light), while $\tfrac{|\f|}{r^3} \gg \tfrac{1}{(r^3)^2}$. This implies that the scalar masses $M_{k,l,n}$ and $\mu_{p,q}$ (and exact one-form spectrum) remain finite, not light, even if $r^3$ is large. Let us check the other modes: the rest of the spectrum is given by the eigenvalues $Y^{k,l,n}_{\pm}$ and $Y^{p,q}_{\pm}$. For large enough $|\f|$, given $|k|$ and $n$, $Y_{\epsilon}^{k,l,n}$ are given by
\beq
Y_{\epsilon}^{k,l,n}= (1 + \epsilon) \frac{1}{2} \f^2 + M_{k,l,n}^2 + 2|\lambda| + \epsilon (2n+3) |\lambda|  + O\left(\frac{1}{\f^2 }\right) \ .
\eeq
We deduce that $Y_{+}^{k,l,n}$ are becoming very large with $r^3$, while $Y_{-}^{k,l,n}$ are staying finite and not light. Similarly, $Y^{p,q}_{+} \sim \f^2 $ become very large. So far, this means there is no mode becoming light in this limit, which is rather unusual. This however happens with $Y^{p,q}_{-}$ in a non-trivial way: one obtains
\beq
Y^{p,q}_{-}= \frac{1}{4} \frac{(P^2 + Q^2)^2}{\f^2} + O\left(\frac{1}{\f^4 }\right) \ .
\eeq
This whole tower of modes becomes light in that limit. With the initial point $\phi_0$ being the one of \eqref{approx}, the mass of these modes was there $m_0 = \sqrt{Y^{p,q}_{-}} \sim \sqrt{P^2 + Q^2}$. The mass is now
\beq
m\ \ \underset{r^3 \rightarrow \infty}{\sim}\ \ m_0 \times \frac{1}{2} \frac{\sqrt{P^2 + Q^2}}{|\f|} = m_0 \times \frac{\pi}{|N|} \sqrt{(p r^2)^2 + (q r^1)^2}\, \frac{1}{r^3} \ .
\eeq
Whether this matches the conjectured behavior \eqref{conj} now depends on the field space distance in $r^3$: it would work for a distance of the form $M_p \ln \left( \tfrac{r^3}{r^3_0} \right) $. Determining this distance however requires to determine the kinetic term for the $r^3$ modulus, which is part of the effective theory. This goes beyond the scope of this work, as discussed in Section \ref{sec:effth}.

Last but not least, there could be an interesting regime, different from the initial one \eqref{approx}, where no mode becomes light: the regime where we change $r^3$ towards $r^3 \sim r^1 \sim r^2$ (one could allow a factor $|N|$, it does not change the discussion). Indeed, none of the Laplacian eigenvalues vanishes by setting $r^3$ to a finite value, they are then all of order $1/(r^m)^2$ times a combination of integers, $2\pi$ and square roots. For instance, considering again the tower $m = \sqrt{Y^{p,q}_{-}}$ with $r^1=r^2$, $(2\pi)^2(p^2+q^2)=X^2$, one obtains
\beq
m\ \ \underset{r^3 \sim r^1}{\sim}\ \ m_0 \times \left(1 + \frac{1}{2} \left(\frac{N}{X}\right)^2  \left(\frac{r^3}{r^1}\right)^2 - \left(\frac{r^3}{r^1}\right)^2 \sqrt{\left(1+ \frac{1}{2} \left(\frac{N}{X}\right)^2 \right)^2-1}  \right)^{\frac{1}{2}} \ ,
\eeq
which is of the order of $m_0$. It is also the case of $\f^2$ which is not light anymore as in \eqref{frhierarchy}, compared to the Kaluza--Klein scale. The effective theory may then be changed, not by the addition of (infinitely many) light states, but rather by the loss of some of them. However, the conjectured behavior \eqref{conj} is not verified. This is interesting if going from \eqref{approx}, namely $r^3 \leq |N| r^3 \ll r^{1,2}$, to $r^3 \sim r^1 \sim r^2$ is at least a Planckian distance in field space: this would be in tension with the refined swampland distance conjecture.

We summarize our findings schematically as follows:
\beq
\hspace{-0.2in} \xymatrix @!0 @R=6.5mm @C=1cm{\mbox{points in field space:}\quad & & & \overset{r^3\, \ll \, r^{1,2}}{\bullet} \ar@{<->}[rrrr]^{{\small\mbox{distance}\, \mbox{?}}}  & & & & \overset{r^3\, \sim \, r^{1,2}}{\bullet} \ar@{<->}[rrrr]^{{\small\mbox{distance}\, \mbox{?}}} & & & & \overset{r^3\, \gg \ r^{1,2}}{\bullet} \\
\mbox{number of light modes:}\qquad & & & \mbox{finite} & & & & \mbox{finite} & & & & \mbox{infinite}  \\
\mbox{masses of light modes:}\qquad & & & 0,\ |\f| & & & & 0 & & & & 0,\ \sqrt{Y_-^{p,q}} }
\eeq
Even though this discussion provides an interesting intuition, we recall that one limitation is to have considered the Laplacian spectrum, instead of the effective theory mass spectrum; the two may differ, as we now explain in Section \ref{sec:effth}. Determining the latter will amount to find masses of fluctuations around a solution. Testing the above conjecture will then require in practice to have $\phi_0$ as being the solution point. In such a framework, it would be interesting to verify whether the above behaviour still holds.

\section{Discussion: the effective theory on the nilmanifold}\label{sec:effth}

In this paper we have determined the complete form spectrum of the  Laplacian  on the three-dimensional Heisenberg nilmanifold: the eigenforms and eigenvalues are summarized in Section \ref{sec:higherforms}. Having this spectrum is a first useful step towards obtaining an effective four-dimensional theory out of a ten-dimensional one. Indeed, it provides a natural expansion basis for the ten-dimensional fields. Obtaining a four-dimensional theory then amounts to truncating the ten-dimensional degrees of freedom to a finite set, potentially corresponding to a subset of the Laplacian eigenmodes.
In Section \ref{sec:light} we obtained such a finite set, built out of the Maurer--Cartan forms, by truncating to the light Laplacian eigenspectrum in the limit of small fiber or large base. Let us first comment on such a truncation.

Truncating to the set of Maurer--Cartan forms on a group manifold (or more generally to left-invariant forms on a coset), i.e.~expanding all ten-dimensional fields on this finite set of forms with constant coefficients, is a well-known reduction ansatz: it goes back to the work by Scherk and Schwarz \cite{Scherk:1979zr} (see \cite{MuellerHoissen:1987cq} for cosets). This type of reduction has been identified as giving a (four-dimensional) gauged supergravity,\footnote{A Scherk-Schwarz reduction and matching to a four-dimensional gauged supergravity has been performed explicitly in the case of ten-dimensional heterotic string in \cite{Kaloper:1999yr}, and for ten-dimensional type IIA supergravity with $D_6/O_6$ in \cite{Villadoro:2005cu, Caviezel:2008ik, DallAgata:2009wsi}; see also \cite{Andrianopoli:2005jv} on the relation between the two, and \cite{Cassani:2012pj} for an example in eleven dimensions.} the gaugings or embedding tensor components corresponding here to the group manifold structure constants $f^a{}_{bc}$. Reviews of gauged supergravities can be found in \cite{Samtleben:2008pe, Trigiante:2016mnt}. Such a reduction is in addition expected to be a CT, i.e.~all solutions of the reduced theory lift to solutions of the ten-dimensional one. One reason is that the exterior derivative only maps the finite set of forms to itself, giving therefore a closed set of modes. A related reason is that the internal coordinate dependence effectively disappears, as both the form coefficients and the $f^a{}_{bc}$ are constant.

In Section \ref{sec:truncation}, we reached a similar result, taking however a different path. Making use of the explicit Laplacian eigenforms, we have built a finite set of forms that would satisfy a list of constraints, summarized in \cite{KashaniPoor:2006si}, meant for reductions on manifolds with SU(3) structure. These constraints are required so that the expansion of ten-dimensional fields on the finite set of modes leads to a four-dimensional ${\cal N}=2$ gauged supergravity: see e.g.~\cite{Grana:2005ny}. As it turns out, the set we have built is again made of certain Maurer--Cartan forms, even though a much bigger set of eigenmodes was shown to verify a subset of the constraints. Our result provides an explicit example where all constraints of \cite{KashaniPoor:2006si} are satisfied, the only other example being that of \cite{KashaniPoor:2007tr}.

Coming back to the truncation of Section \ref{sec:light}, the result there is that, for the first time to our knowledge, the Scherk-Schwarz reduction ansatz or truncation is derived as a {\it low energy approximation} of the spectrum of the Laplacian: indeed, while one could in full generality develop all ten-dimensional fields on the basis of eigenforms of the Laplacian, we have shown that a low energy approximation would restrict this expansion to a finite set of modes \eqref{lightmodes}, corresponding precisely to the Scherk-Schwarz reduction ansatz, i.e.~left-invariant forms with constant coefficients.\footnote{In \cite{Caviezel:2008ik} it was argued that a similar result  holds for $\mmm$  being the Iwasawa manifold, however the complete spectrum of the Laplacian on $\mmm$ was not computed therein.}

One may then wonder whether the resulting four-dimensional gauged supergravity is a LEEA. In general, this point has not been settled beyond the case of a Calabi-Yau manifold without flux: see e.g.~related discussions in \cite{Grana:2005ny, Grana:2013ila}. However, it is often (implicitly or explicitly) assumed to hold, and various four-dimensional gauged supergravities are then used for string phenomenology. Unfortunately the answer cannot be settled here: even though one develops the fields on a basis of light modes, two further phenomena could prevent one from getting a low energy effective theory. First, additional ten-dimensional fluxes could bring about different energy scales that could complicate or violate the approximations. Second, a conspiracy in the mass matrix of the theory could make certain linear combinations of heavy modes, i.e.~truncated ones, become  accidentally light. Such modes are sometimes called ``space invaders'' \cite{Duff:1986hr}; see \cite{Anderson:2014yva} for a more recent example. This shows that  the usual requirement of having flux scales (or ``moduli masses'') much smaller than the Kaluza-Klein scales, as here in \eqref{frhierarchy}, is strictly-speaking not sufficient.

The way to determine a low energy effective theory is to consider (linear) fluctuations around a given (ten-dimensional) solution, and study the resulting  mass matrix thus identifying all potentially light modes \cite{Duff:1986hr}. This procedure is the one that takes ``space invaders'' into account. It is the same procedure that identifies harmonic forms as the correct finite set of light (massless) modes on a Calabi-Yau manifold without flux. We expect the spectrum of Laplacian eigenforms determined here to be very useful in this task, providing a natural expansion basis for all fluctuations around a solution containing the Heisenberg nilmanifold. The low-energy approximation and truncation proposed in Section \ref{sec:light} could serve as a guide for a low-energy approximation of the complete theory. Obtaining such a LEEA would be useful in view of our discussion on the refined swampland distance conjecture in Section \ref{sec:swamp}. Thanks to the kinetic terms, one could calculate distances in field space. This would allow to determine whether there is a tension with the conjecture, as we pointed out. We hope to return to this program in future work.

\vspace{0.4in}

\subsection*{Acknowledgements}

We would like to thank D.~Cassani, A.-K.~Kashani-Poor, F.~Marchesano, R.~Minasian, H.~Samtleben, M.~Trigiante and T.~Weigand for very useful discussions.

\end{document}